\documentclass[twocolumn,superscriptaddress,prb,aps,longbibliography]{revtex4}
\usepackage{amsxtra,amssymb,bm,amsbsy,amsmath,amsthm,latexsym,dsfont,array,layout,graphicx,physics,mathrsfs,color,soul,subfigure,enumerate,footmisc,scrextend}
\usepackage[colorlinks=true,citecolor=blue,linkcolor=blue,anchorcolor=blue,urlcolor=blue]{hyperref}
\hypersetup{linktoc=all}

\begin{document}

\title{Generating Majorana qubit coherence in Majorana Aharonov-Bohm interferometer}
\author{Fei-Lei Xiong}
\thanks{These two authors contributed equally to this work.}
\affiliation{Department of Physics and Center for Quantum Information Science,
National Cheng Kung University, Tainan 70101, Taiwan}
\author{Hon-Lam Lai}
\thanks{These two authors contributed equally to this work.}
\affiliation{Department of Physics and Center for Quantum Information Science,
National Cheng Kung University, Tainan 70101, Taiwan}
\author{Wei-Min Zhang}
\email{wzhang@mail.ncku.edu.tw}
\affiliation{Department of Physics and Center for Quantum Information Science,
National Cheng Kung University, Tainan 70101, Taiwan}
\affiliation{Physics Division, National Center for Theoretical Sciences, Taipei 10617, Taiwan}

\begin{abstract}

We propose an Aharonov-Bohm interferometer consisted of two topological superconducting chains (TSCs) to generate coherence of Majorana qubits, each qubit is made of two Majorana zero modes (MZMs) with the definite fermion parity. We obtain the generalized exact master equation as well as its solution and study the real-time dynamics of the MZM qubit states under various operations. We demonstrate that by tuning the magnetic flux, the decoherence rates can be modified significantly, and dissipationless MZMs can be generated. By applying the bias voltage to the leads, one can manipulate MZM qubit coherence and generate a nearly pure superposition state of Majorana qubit. Moreover, parity flipping between MZM qubits with different fermion parities can be realized by controlling the coupling between the leads and the TSCs through gate voltages. 
\end{abstract}

\maketitle

\section{Introduction}

Topological quantum computation has been widely investigated as a promising candidate for realizing fault-tolerant quantum 
computation due to its robustness against decoherence~\cite{K01,NSS08}.  The protection against decoherence during the 
computation process relies on highly-degenerate ground states of the qubit space, which is realized by spatially separated 
Majorana zero modes (MZMs)~\cite{K01}. Theories have predicted that under certain physical conditions, MZMs can exist 
at the ends of 1D effective spinless $p$-wave superconductors~\cite{K01}, which can be generated by contacting a conventional 
$s$-wave superconductor to topological insulators~\cite{FK08,FK09,CF11,SZH16}, magnetic atom chains~\cite{BS13,PG13,HK14,SI14,RP15,PK16}, 
or semiconductors with strong spin-orbit interaction~\cite{SL10,A10,LSD10,ORV10,DB11,CZ11,PL12,LBK18}.
After nearly a decade of effort, experimentalists have recently observed some signatures of MZMs in the proximitized 
nanowire systems~\cite{MZF12,DRM12,FVM13,AHA16,DVH16,CYS17,NDW17,SKH17,GZB18}. 

Because of the non-Abelian exchange property of MZMs, braidings among them correspond 
to nontrivial unitary transformations, which plays a central role in the scheme of topological 
quantum computation~\cite{NSS08}. 
In the literature, there are mainly two kinds of methods to realize the braiding operations, 
either by changing the physical parameters of the system adiabatically~\cite{LZ20,AO11,SC11,HV13,AH16} 
or by performing projective measurements systematically~\cite{BF08,BF09,PR17,KK17}. 
For instance, MZMs can be moved along nanowires by tuning the chemical potential and 
braiding operations can be performed in T-junction structures~\cite{AO11,LZ20}. Other proposals 
include performing braidings by controlling the couplings between different MZMs~\cite{SC11}, 
or by fusions of different MZMs in T-junction nanowires~\cite{AH16}. As for the measurement-based 
braiding method, effective braidings of two MZMs are done by measuring the joint fermion parity of 
the MZM pair rather than exchanging their spatial positions. This kind of measurements, as well as 
the error correction code, can be realized by coupling quantum dots to MZMs~\cite{KK17}, or by 
using Aharonov-Bohm interferometers~\cite{LBK18,LPS16,PLS16,AHM16,PR17}. 

\begin{figure}[]
	\centering
	\includegraphics[width=0.48\textwidth]{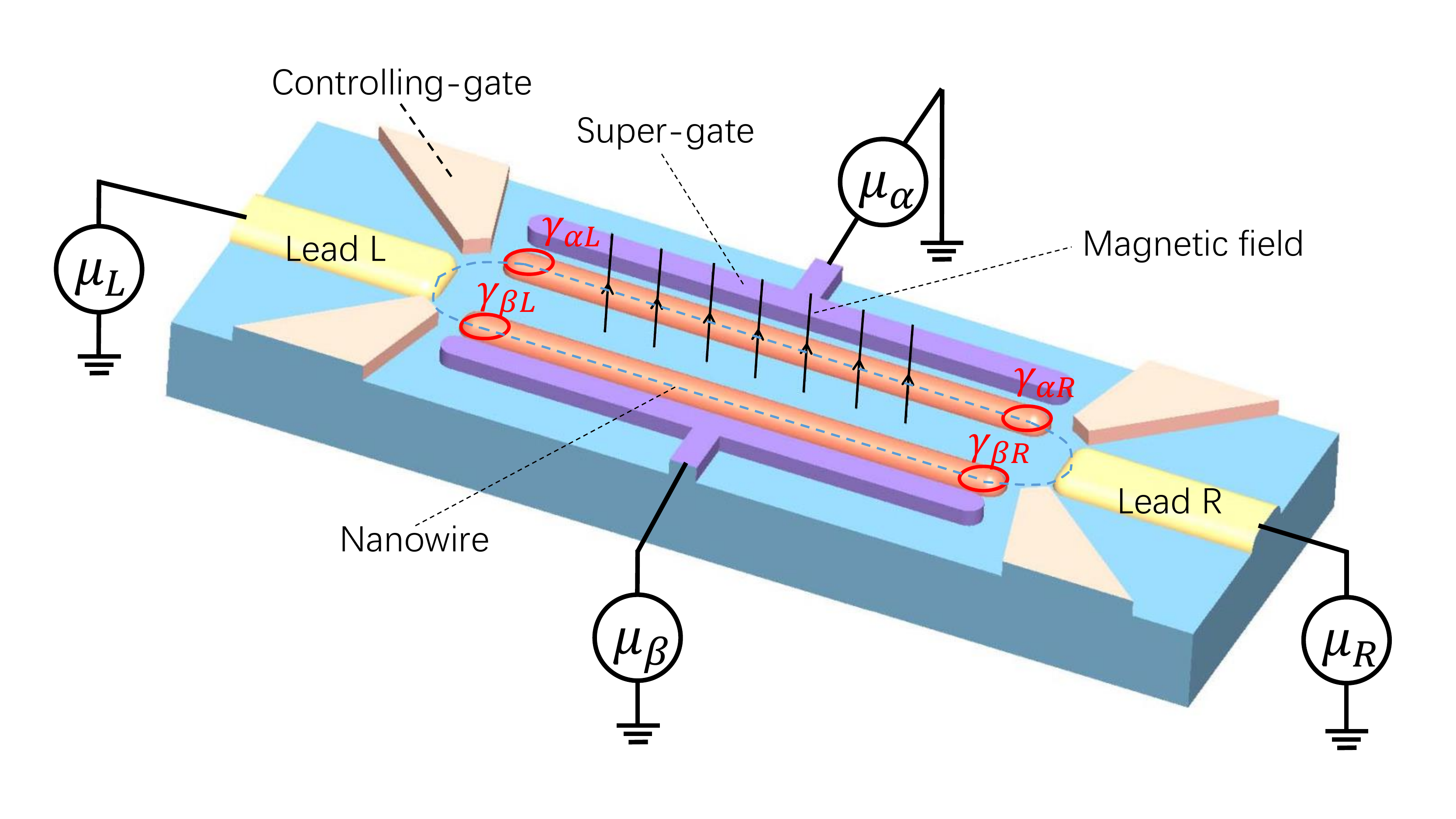}
	\caption
	{ \small (Color Online) The schematic picture of the proposed Majorana AB interferometer. The two 
	copper-colored wires in the picture are the proximitized TSCs. By tuning the chemical potential $\mu_\alpha$ 
	and $\mu_\beta$ with the super-gates, two MZMs can be formed at the ends of each TSC (See the 
	red circles in the figure). The leads L and R are two electrodes. In the central region between the TSCs, 
	we apply a magnetic flux threading into the central region of the interferometer. 
	The controlling gates are used for controlling the tunneling amplitudes.}
	\label{fig1}
\end{figure}

However, in realistic situations, the Majorana qubits are unavoidably coupled to external controlling 
gates under qubit operations~\cite{GC11,BW12,HYZ20,LZ20}. As a consequence, the topological 
protection against decoherence can be destroyed by, for instance, charge fluctuations of the controlling 
gates~\cite{SR12,LY18,LZ20}. In this paper, we propose a new scheme other than braidings to manipulate 
the qubit states of MZMs, where noise effects are taken into account. 
Our device mainly consists of an Aharonov-Bohm (AB) interferometer, 
which is constructed by connecting two TSCs with two metal leads (See Fig.~\ref{fig1}). The TSCs are 
tuned to the topological phase through the super-gates so that four MZMs are formed at their ends. 
Two leads with tunable bias voltage are coupled to the left and right ends of the TSCs, with the coupling 
strengths being also adjustable through the controlling gates. The magnetic flux 
threading into the central region of the interferometer can affect the interference pattern of the interferometer. 

In this device, unlike the braiding operations, the fermion parity of the MZM states can be intentionally switched by 
letting electrons tunnel into and out of the TSCs from the leads. In addition, electrons transport coherently from 
one end of a TSC to the other through the MZM pairs, the two paths formed by the TSCs interfere with each other. 
Quantum coherence can therefore be generated and controlled by tuning the lead-TSC couplings through gate 
voltages or by the applied magnetic flux. Through the exact master equation involving pairing 
interaction~\cite{LY18,LZ20,HYZ20,Z19}, we shall study the real-time dynamics of this Majorana AB interferometer 
under various operations. We discover that under the magnetic flux controlling, dissipationless MZM modes can be 
formed between the leads and the TSCs. Also, intended MZM qubit states  
with different parities can be prepared by applying a bias to the leads. Finally, parity flipping between the  
MZM qubits with different fermion parity can 
be done by controlling the coupling strengths between the leads and the TSCs through gate voltages.

Our paper is organized as follows. In Sec.~\ref{sec_2}, we propose the model of the Majorana AB interferometer, which includes two TSCs 
contacting with left and right leads. The magnetic flux threading into the central region of the interferometer. We construct the Hamiltonian 
of the electron tunnelings in this superconducting AB interferometer incorporating the magnetic flux.  
In Sec.~\ref{sec_3}, we derive the 
exact master equation for the MZMs localized at the ends of two TSCs. We show that the damping of the 
MZMs and the couplings induced by the leads are explicitly 
related to the generalized non-equilibrium Green functions for the MZMs. Furthermore, the density matrix of the four MZMs 
(two MZM qubits with different parity respectively) can be obtained as the solution to our 
exact master equation. In Sec.~\ref{sec_4}, we discuss Majorana qubit state evolution under various kinds of operations. We show that 
dissipationless MZM modes will be formed by controlling the magnetic flux. Moreover, MZM qubit coherence can be generated 
and controlled when a bias is applied between the two leads. The MZM qubit state parity can also be flipped by 
tuning the couplings between the MZMs and the leads. Finally, the conclusions are summarized in Sec.~\ref{sec_5}. 

\section{The Majorana AB interferometer and its modeling}
\label{sec_2}
The Majorana AB interferometer we propose is schematically plotted in Fig.~\ref{fig1}. The nanowires labeled 
$\alpha$ and $\beta$ are two 1D spinless $p$-wave superconductor chains, which can be realized by, for example, 
strong spin-orbit interacting nanowires proximitized by conventional $s$-wave superconductors. In this paper, 
we model them as $N$-site Kitaev chains and in the absence of the magnetic flux, they are described by the Hamiltonians~\cite{K01}
\begin{align}
H_{\alpha(\beta)}\!=\!&\sum^{N-1}_{j=1}\big[\!-\!w a_{\alpha(\beta),j}^{\dagger}a_{\alpha(\beta),j+1} \nonumber\\
\!&+\!{\Delta} e^{{i\phi_{\alpha(\beta)}}} a_{\alpha(\beta),j} a_{\alpha(\beta),j+1} \!+\! h.c.\big] \nonumber\\
&-\sum^{N}_{j=1}\mu_{\alpha(\beta)}\big[a_{\alpha(\beta),j}^\dagger a_{\alpha(\beta),j}-\frac{1}{2}\big] .
\label{H_Kitaev}
\end{align}
Here, 
$a_{\alpha,j}$ ($a_{\alpha,j}^\dagger$) denotes the annihilation (creation) operator of site-$j$ in chain 
$\alpha$ (similarly for $\beta$). The hopping amplitude $w$ is real-valued and $\Delta e^{i\phi_{\alpha(\beta)}}$, 
with $\Delta$ and $\phi_{\alpha(\beta)}$ being real numbers, are the superconducting gap in chain $\alpha$ and $\beta$. 
Two leads, which are labeled L and R and modelled by the free electron gas Hamiltonians
\begin{align}
H_{L(R)}=\sum_k \epsilon_{L(R)k} c_{L(R)k}^\dagger c_{L(R)k}\,,
\end{align}
are coupled to the TSCs through the tunneling Hamiltonian
\begin{align}
H_I=&\sum _k\big(\lambda_{{\alpha Lk}}(t) c_{{Lk}}^{\dagger } a_{\alpha ,1} \!+\!\lambda_{{\beta Lk}} (t) c_{{Lk}}^{\dagger }  a_{\beta ,1}\nonumber\\
&\!+\!\lambda_{{\alpha Rk}} (t) c_{{Rk}}^{\dagger } a_{\alpha ,N}
\!+\!\lambda_{{\beta Rk}} (t) c_{{Rk}}^{\dagger } a_{\beta ,N}\!+\!h.c.\big)\,.
\end{align} 
Here, $\epsilon_{L(R) k}$ is the single-particle energy of mode-$k$ in lead L(R), with $c_{L(R)k}$ and $c^\dagger_{L(R)k}$ 
being the corresponding annihilation and creation operators respectively. Moreover, $\lambda_{\alpha L k}(t)$, 
$\lambda_{\beta L k}(t)$, $\lambda_{\alpha R k}(t)$, and $\lambda_{\beta R k}(t)$ stand for the coupling strengths 
between the modes in the leads and the ends of the TSCs. They are controlled by tuning the controlling gates 
in Fig.~\ref{fig1} and can in general be time-dependent. In the following, if not specified, we would omit the 
time-dependence of the $\lambda$'s.  

In addition, we apply the magnetic flux threading into the central region of the interferometer $\Phi\!=\!\oint \bm{A}(\bm{r}) 
\cdot d\bm{r}$, where $\bm{A}(\bm{r})$ stands for the vector potential at position $\bm{r}$. 
The operators $a_{\alpha,j}$ and $a_{\beta,j}$ in the Hamiltonians $H_\alpha$ and $H_\beta$ should be converted 
according to Peierls substitution
\begin{align}
a_{\alpha(\beta),j}\rightarrow e^{-i \phi_{\alpha(\beta),j}} a_{\alpha(\beta),j}\,,
\end{align}
and the phase functions satisfy the relation
\begin{align}
\phi_{\alpha(\beta),j+1}-\phi_{\alpha(\beta),j}=\frac{e}{\hbar}\int_{\alpha(\beta),j}^{\alpha(\beta),j+1} \bm{A}(\bm{r}) \cdot d\bm{r}\,.
\end{align}
Therefore, the Hamiltonians of the TSCs $\alpha$ and $\beta$ 
with magnetic flux threading into the central region of the interferometer can be written as
\begin{align}\label{eq_Hab}
H_{\alpha(\beta)} & \!=\!\sum_j\big[\!-\!w e^{\!-\!i[{\phi}_{\alpha(\beta) ,j+1}\!-\!{\phi}_{\alpha(\beta),j}]}  a_{\alpha(\beta),j}^{\dagger}  a_{\alpha(\beta),j+1}  \nonumber\\
&\!+\!{\Delta} e^{{i\phi_{\alpha(\beta)}}} e^{\!-\!i[{\phi}_{\alpha(\beta),j+1}\!+\!{\phi}_{\alpha(\beta),j}]} a_{\alpha(\beta),j} a_{\alpha(\beta),j+1} \!+\! h.c.\big]\nonumber\\
&-\sum_j\mu_{\alpha(\beta)}\big[a_{\alpha(\beta),j}^\dagger a_{\alpha(\beta),j}-\frac{1}{2}\big]\,.
\end{align}
Apply the substitutions that 
\begin{align}
\tilde{a}_{\alpha(\beta),j}=e^{i\phi_{\alpha(\beta)}/2} e^{-i\phi_{\alpha(\beta),j}}a_{\alpha(\beta),j}\,, 
\end{align}
the Hamiltonians $H_\alpha$, $H_\beta$ and $H_I$ can be expressed in terms of the operators $\tilde{a}_{\alpha,j}$ and $\tilde{a}_{\beta,j}$, i.e.,
\begin{align}\label{eq_A_H}
H&_{\alpha(\beta)} \nonumber\\
&=\!\!\sum_j\!\Big[\!-\!w \tilde{a}_{\alpha(\beta),j}^\dagger \tilde{a}_{\alpha(\beta),j+1}+\Delta \tilde{a}_{\alpha(\beta),j} \tilde{a}_{\alpha(\beta),j+1}+h.c. \Big]\nonumber\\
& -\sum_j\mu_{\alpha(\beta)}\Big[\tilde{a}_{\alpha(\beta),j}^\dagger \tilde{a}_{\alpha(\beta),j}-\frac{1}{2}\Big]\,,\\
H&_{I}\!=\!\sum_k (\lambda_{\alpha L k} 
e^{i \tilde{\phi }_{\alpha L}} c_{L k}^\dagger \tilde{a}_{\alpha,1} \!+\!\lambda_{\beta L k} e^{i \tilde{\phi}_{\beta L}} c_{Lk}^\dagger \tilde{a}_{\beta,1}\nonumber\\&
\!+\!\lambda_{\alpha R k} 
e^{i \tilde{\phi }_{\alpha L}} c^\dagger_{Rk} \tilde{a}_{\alpha,N} \!+\!\lambda_{\beta R k} e^{i \tilde{\phi }_{\beta R} } c^\dagger_{Rk} \tilde{a}_{\beta,N}\!+\!h.c.)\,. \label{eq_A_HT}
\end{align}
where $\tilde{\phi}_{\alpha(\beta)L}\!=\!\phi_{\alpha(\beta),1}-{\phi_{\alpha(\beta)}}/{2}$ and $\tilde{\phi}_{\alpha(\beta)R}\!=\! \phi _{\alpha(\beta),N}-{\phi_{\alpha(\beta)}}/{2}$. 

The nanowire chemical potentials $\mu_\alpha$ and $\mu_\beta$, the hopping amplitude $w$, and the pairing parameter $\Delta$ 
are tuned so that two MZMs can be generated at the ends of each nanowire, which we denote as 
$\gamma_{\alpha L}$, $\gamma_{\alpha R}$, $\gamma_{\beta L}$ and $\gamma_{\beta R}$, respectively (See the red circles 
in the ends of two TSCs in Fig.~\ref{fig1}). To illustrate this property, we consider an ideal parameter setting that $\mu_\alpha
=\mu_\beta=0$ and $\Delta\!=\!w$, then the MZMs possess the explicit form,~\cite{K01}
\begin{subequations}\label{eq_gamma_def}
\begin{alignat}{2}
&\gamma_{\alpha L}\!=\!\tilde{a}_{\alpha ,1}\!+\!\tilde{a}_{\alpha ,1}^{\dagger }\,, \quad
&&\gamma_{\alpha R}\!=\!-i \tilde{a}_{\alpha ,N}\!+\!i \tilde{a}_{\alpha ,N}^{\dagger }\,, \quad  \\
&\gamma_{\beta L}\!=\!\tilde{a}_{\beta ,1}\!+\! \tilde{a}_{\beta ,1}^{\dagger }\,, \quad
&&\gamma_{\beta R}\!=\!-i \tilde{a}_{\beta ,N}\!+\!i \tilde{a}_{\beta ,N}^{\dagger }\,,
\end{alignat}
\end{subequations}
and the annihilation operators of the zero-energy quasiparticle excitations are 
\begin{align}
b_{\alpha,0}\!=\!\frac{1}{2} (\gamma_{\alpha R}\!+\!i \gamma_{\alpha L})\,, \quad
b_{\beta,0}\!=\!\frac{1}{2} (\gamma_{\beta R}\!+\!i \gamma_{\beta L})\,.
\end{align}  
In this work, we consider the case that the bias between the chains and the leads is much smaller than the superconducting 
gap and the excitation of quasiparticles in the continuous bands of the TSCs is negligible~\cite{BD07}. As a consequence, 
in the interaction Hamiltonian~\eqref{eq_A_HT}, the components of the field operators that involved with the non-zero energy 
bogoliubons in the TSCs can be neglected, 
Then the interaction Hamiltonian is reduced to
\begin{align}
&H_I\!=\!\frac{1}{2}\sum_k(\lambda_{{\alpha Lk}}  e^{i \tilde{\phi}_{{\alpha L}}} c_{{Lk}}^{\dagger } \gamma_{{\alpha L}}
\!+\! \lambda_{{\beta Lk}}  e^{i \tilde{\phi}_{{\beta L}}} c_{{Lk}}^{\dagger }\gamma_{{\beta L}} \nonumber\\
&\!+\!i  \lambda_{{\alpha Rk}} e^{{i\tilde{\phi} }_{{\alpha R}}}  c_{{Rk}}^{\dagger } \gamma_{{\alpha R}} \!+\!i  \lambda_{{\beta Rk}} 
e^{{i\tilde{\phi} }_{{\beta R}}} c_{{Rk}}^{\dagger } \gamma_{{\beta R}} \!+\!h.c.)\,.\label{eq_HI}
\end{align}
In Eq.~\eqref{eq_Hab}, the pairing phases at the left and right ends of chain-$\alpha$ and the left and right ends of chain-$\beta$ 
are $-2\tilde{\phi }_{\alpha L}$, $-2\tilde{\phi }_{\alpha R}$, $-2\tilde{\phi }_{\beta L}$ and $-2\tilde{\phi }_{\beta R}$, respectively. 
Following the convention  in Feynman's dealing with the pairing phases in the Josephson junctions~\cite{F11}, the phases satisfy 
\begin{subequations}\label{eq_delta}
\begin{alignat}{2}
&\tilde{\phi}_{\alpha L}-\tilde{\phi}_{\beta L}&&=-\frac{\phi}{2}-\pi \frac{\Phi}{\Phi_0}\,,\label{eq_deltaL}\\
&\tilde{\phi}_{\alpha R}-\tilde{\phi}_{\beta R}&&=-\frac{\phi}{2}+\pi \frac{\Phi}{\Phi_0}\,,\label{eq_deltaR}
\end{alignat}
\end{subequations}
where $\Phi_0=h/e$ is the flux quantum with $h$ standing for the Planck constant and $e$ standing for the elementary charge, 
$\phi=\phi_{\alpha}-\phi_{\beta}$ is the initial pairing phase difference between the two TSCs. 

Although our modeling of the Majorana AB interferometer is based on modeling the TSCs as Kitaev chains under special 
physical conditions, it is applicable to more general cases. Generally speaking, when the effective 1D spinless $p$-wave 
superconductors are in the topological phase and the excitation of quasiparticles in the continuous band is negligible, 
the Hamiltonians \eqref{H_Kitaev} characterizing the TSCs can be written as $H_{\alpha}=i\epsilon_{\alpha}\gamma_{\alpha L}\gamma_{\alpha R}$ 
and $H_{\beta}=i\epsilon_{\beta}\gamma_{\beta L}\gamma_{\beta R}$, where the $\gamma$'s are 
the Majorana operators with wave packets localized near the ends of the TSCs~\cite{K01}. 
The energy $\epsilon_{\alpha(\beta)}\sim 0$ is proportional to the wavefunction overlap of the MZMs $\gamma_{\alpha(\beta)L}$ 
and $\gamma_{\alpha(\beta)R}$, which is exponentially suppressed by the length of the TSCs. If the TSCs are long enough, 
the wave packets of the $\gamma$'s can be seen as localized and the energy values $\epsilon_{\alpha(\beta)}$ can be treated 
as zero. The pairing phase differences between the ends of the TSCs can also apply to the relations in Eq.~\eqref{eq_delta}. 
As a consequence, Eq.~\eqref{eq_HI} and Eq.~\eqref{eq_delta} together describe the interaction between the TSCs and the 
leads, except for the fact that the Majorana operators are no longer in the form of Eq.~\eqref{eq_gamma_def}. Thus,
the leads and the TSCs together form an Aharonov-Bohm interference ring. Particles exchange and interfere through the MZMs 
in the TSCs. The  dynamics of the system is influenced by the magnetic flux $\Phi$ and the time-dependent tunneling amplitudes, 
through which we can generate coherence between the two TSCs and manipulate the MZM qubit states.  

\section{The exact master equation and the density matrix}
\label{sec_3}

\subsection{The exact master equation}
We treat the MZMs as the principal system, and the two leads as the environment. Suppose that the total system is initially in a 
product state $\rho_{\rm{tot}}(0)=\rho(0)\otimes \rho_L(0)\otimes \rho_R(0)$, where $\rho(0)$ is the state of the principal system 
and $\rho_L(0)$ ($\rho_R(0)$) is the state of lead L (R). Without loss of generality, we also assume that  $\rho_L(0)$ ($\rho_R(0)$) 
is the thermal equilibrium state associated to temperature $T_L$ ($T_R$) and chemical potential $\mu_L$ ($\mu_R$). By taking 
advantage of the path integral approach in the coherent state representation, states of the system can be found to evolve according 
to the exact master equation~\cite{TZ08,LZ12,HYZ20,Z19}
\begin{align}\label{eq_rho}
\dot{\rho}(t)&\!=\!
-\frac{1}{i \hbar}[\rho(t), \tilde{H}_L(t)\!+\!\tilde{H}_R(t)]\nonumber\\
&+\!\sum_{i,j} \frac{\Gamma_{Lij} (t)}{2} \big[ \gamma_{iL} \rho(t) \gamma_{jL}\!-\!\frac{1}{2}\{\rho(t), \gamma_{jL}\gamma_{iL}\}\big]\nonumber\\
&+\!\sum_{i,j} \frac{\Gamma_{Rij} (t)}{2} \big[ \gamma_{iR} \rho(t) \gamma_{jR}\!-\!\frac{1}{2}\{\rho(t), \gamma_{jR}\gamma_{iR}\}\big]\,.
\end{align}
In the formula, 
\begin{align}\label{eq_HL}
 \tilde{H}_{L}(t)=\frac{i}{4}([\dot{\bm{U}}_{L}{\bm{U}}_{L}^{-1}]_{\alpha\beta}-[\dot{\bm{U}}_{L}{\bm{U}}_{L}^{-1}]_{\beta\alpha}) \gamma_{\alpha L} \gamma_{\beta L} 
\end{align}
is the environment-induced renormalized Hamiltonian for the left-side MZMs; $\Gamma_{Lij} (t)$ ($i,j=\alpha,\beta$) characterize 
the decoherence rates of the MZMs in the left side and can be explicitly written in terms of the generalized non-equilibrium 
Green functions $\bm{U}_L$ and $\bm{V}_L$,
\begin{align}\label{eq_gammaL}
\Gamma_{Lij}(t)=[\dot{\bm{V}}_{L}-(\dot{\bm{U}}_{L} {\bm{U}}_{L}^{-1} \bm{V}_{L} +h.c.)]_{ij}\,.
\end{align}
In Eqs.~\eqref{eq_HL}-\eqref{eq_gammaL}, $\bm{U}_L$ and $\bm{V}_L$ are short for the retarded Green's function $\bm{U}_L(t,t_0)$
 and the correlation function $\bm{V}_L(t)$ involving pairing interactions~\cite{XZ20}. $\bm{U}_L(t,t_0)$ 
satisfies the integro-differential equation 
\begin{align}
\partial_t \bm{U}_L(t,t_0)+2\int_{t_0}^t d\tau \bm{g}_L(t,\tau) \bm{U}_L(\tau,t_0)=0\,,
\end{align} 
with the initial condition that $\bm{U}_L(t_0,t_0)\!=\!\bm{I}$ ($\bm{I}$ is a $2\times 2$ identity matrix). The time-nonlocal integral kernel $\bm{g}_L(t,\tau)$ is given by 
\begin{align}
&\bm{g}_L(t,\tau)\!=\!\int\! 
\frac{d\epsilon}{2\pi} \big[ e^{-i \epsilon(t-\tau)} \bm{J}_L^e \!+\!e^{i \epsilon(t-\tau)} \bm{J}_L^h \big],
\end{align}
where $\bm{J}_L^e$ and $\bm{J}_L^h$ are short for the electron spectral density function $\bm{J}_L^e(\epsilon,t,\tau)$ and the hole spectral 
density function $\bm{J}_L^h(\epsilon,t,\tau)$, respectively; and $\bm{J}_L^h(\epsilon,t,\tau)=\bm{J}_L^{e*}(\epsilon,t,\tau)$. 
Define that $(\bm{J}_{L}^0)_{ij}\!=\!\frac{\pi}{2}\sum_k \delta(\epsilon\!-\!\epsilon_{L k})\! \lambda_{i Lk}^*(t)\lambda_{j Lk}(\tau)$, where $i$ 
and $j$ are either $\alpha$ or $\beta$, then the complete expression of $\bm{J}_L^e$ is
\begin{align}
&\bm{J}_L^e=
\mqty*(
    [\bm{J}_{L}^0]_{\alpha\alpha}  &  [\bm{J}_{L}^0]_{\alpha\beta} e^{-i \delta_L/2}  \\
    [\bm{J}_{L}^0]_{\beta\alpha} e^{i \delta_L/2}  &  [\bm{J}_{L}^0]_{\beta\beta}
)\!, 
\end{align}
where $\delta_{L}=\tilde{\phi}_{\alpha L}-\tilde{\phi}_{\beta L}$. Note that the cross coupling between $\alpha$ and $\beta$ is dependent on 
$\delta_{L}$ which has been defined in Eq.~\eqref{eq_delta}. 
$\bm{V}_L(t)$  can be written in terms of the retarded Green's function $\bm{U}_L$ that
\begin{align}
\bm{V}_L(t)\!=\!2\int_{t_0}^t{d}\tau_1 \!\int_{t_0}^t{d}\tau_2 \bm{U}_L(t,\tau_1) \tilde{\bm{g}}_L(\tau_1,\tau_2) \bm{U}_L(\tau_2,t)\,,  
\end{align}
where the system-environment correlation $\tilde{\bm{g}}_L(\tau_1,\tau_2)$ satisfies 
\begin{align}
\tilde{\bm{g}}_L(\tau_1,\tau_2)\!=\!\int\! 
\frac{d\epsilon}{2\pi} \Big[ f_L^e(\epsilon) e^{-i \epsilon(\tau_1-\tau_2)} \bm{J}_L^e
\!+\! f_L^h(\epsilon) e^{i \epsilon(\tau_1-\tau_2)} \bm{J}_L^h \Big]\,.
\end{align}
Here, $f_L^e(\epsilon)=\frac{1}{e^{\beta_L(\epsilon-\mu_L)}+1}$ and $f_L^h(\epsilon)=1-f_L^e(\epsilon)$ are the initial particle number 
distribution of electrons and holes in lead L respectively. 
All the relations and conventions are similar for the right-side MZMs, with only the index $L$ being replaced by $R$.  

As shown in Eq.~\eqref{eq_rho}, couplings between the TSCs and the leads induce interactions among the MZMs as well as the dissipation of them. 
Specifically, the MZMs $\gamma_{\alpha L}$ and $\gamma_{\beta L}$ ($\gamma_{\alpha R}$ and $\gamma_{\beta R}$) in the left (right) 
side are coupled to each other through the renormalized Hamiltonian $\tilde{H}_{L}$ ($\tilde{H}_{R}$), and dissipate to lead $L$ ($R$) 
through the dissipation coefficients $\Gamma_{L(R)ij}$. All the MZM dynamics can be captured by the Majorana correlation function matrix 
\begin{align}\label{eq_Mt}
\bm{M} & (t)\!=\notag \\
& \mqty*(
 0 & \langle i \gamma _{{\alpha L}} \gamma _{{\beta L}} \rangle & \langle i \gamma _{{\alpha L}} \gamma _{{\alpha R}} \rangle & \langle i \gamma _{{\alpha L}} \gamma _{{\beta R}} \rangle \\ ~& ~ & ~ & \\
\langle -i \gamma _{{\alpha L}} \gamma _{{\beta L}} \rangle & 0 & \langle i \gamma _{{\alpha R}} \gamma _{{\beta L}} \rangle & \langle i \gamma _{{\beta L}} \gamma _{{\beta R}} \rangle \\~& ~ & ~ & \\
\langle -i \gamma _{{\alpha L}} \gamma _{{\alpha R}} \rangle & \langle -i \gamma _{{\alpha R}} \gamma _{{\beta L}} \rangle & 0 & \langle i \gamma _{{\alpha R}} \gamma _{{\beta R}} \rangle \\~& ~ & ~ & \\
\langle -i \gamma _{{\alpha L}} \gamma _{{\beta R}} \rangle & \langle -i \gamma _{{\beta L}} \gamma _{{\beta R}} \rangle & \langle -i \gamma _{{\alpha R}} \gamma _{{\beta R}} \rangle & 0 
)\!,
\end{align}
which can be obtained in terms of non-equilibrium Green functions, explicitly,
\begin{align}\label{eq_M}
\bm{M}(t)=\bm{U}\bm{M}(t_0)\bm{U}^{T}\! - \!\frac{i}{2}(\bm{V}\! - \!\bm{V}^{T}),
\end{align}
where the superscript T denotes the matrix transpose. Also, the expectation value of the fermion parity~\cite{K01} for the MZM states  can be written as
\begin{align}\label{eq_gamma}
\bar{P}(t)&\!=\!-\big\langle\gamma _{{\alpha L}} \gamma _{{\alpha R}}\gamma _{{\beta L}} \gamma _{{\beta R}}\big\rangle \nonumber\\
&\!=\!\bar{P}(t_0)\det(\bm{U})\! +\!\frac{1}{4}\big(V_{L \alpha \beta }\! -\!V_{L\alpha \beta }^{*}\big)\big(V_{R \alpha \beta }\! -\!V_{R\alpha \beta }^{*}\big)\,.
\end{align}
Note that in Eqs.~\eqref{eq_Mt}-\eqref{eq_gamma}, we have omitted the time-dependence of the Majorana operators. The Green functions  
$\bm{U}$ and $\bm{V}$ can be explicitly expressed as 
$\bm{U}(t,t_0)\!=\!\mqty*(\bm{U}_L &  \bm{0}  
\\  \bm{0}  &  \bm{U}_R )\!$ and $\bm{V}(t)\!=\! \mqty*(\bm{V}_L &  \bm{0}  
\\  \bm{0}  &  \bm{V}_R )$ respectively. 

\subsection{Exact dynamics of the density matrix}

The MZM density matrix in the AB interferometer can be obtained by solving the master equation. In the following, the basis 
$\{\ket{0},b^{\dag}_{0,\alpha}\ket{0},b^{\dag}_{0,\beta}\ket{0},b^{\dag}_{0,\alpha}b^{\dag}_{0,\beta}\ket{0}\}$ is used, which
consists of two Majorana qubit basis with different parities, the even parity qubit basis $\{\ket{0},b^{\dag}_{0,\alpha}b^{\dag}_{0,\beta}\ket{0}\}$ and the old parity 
qubit basis $\{b^{\dag}_{0,\alpha}\ket{0},b^{\dag}_{0,\beta}\ket{0}\}$, where the operator $b^{\dag}_{0,\alpha(\beta)}=
\frac{1}{2}(\gamma_{\alpha(\beta)R}-i\gamma_{\alpha(\beta)L})$ creates a zero-energy Bogoliubon in TSC $\alpha$ ($\beta$). 
This basis corresponds to the zero-energy Bogoliubon occupation in $\alpha$, $\beta$ or both TSCs. We consider the case that initially 
the two nanowires are not correlated, i.e., the system initial state reads
\begin{align}
\rho(t_0)= \mqty*(
\rho_{00}(t_0)  &    0    &    0    & 0 \\
0        & \rho_{\alpha\alpha}(t_0) & 0 & 0 \\
0        & 0 & \rho_{\beta\beta}(t_0) & 0 \\
0  & 0 & 0 & \rho_{dd}(t_0)
),
\end{align}
where the subscripts $0$, $\alpha$, $\beta$, and $d$ correspond to the states $\ket{0}$, $b^{\dag}_{0,\alpha}\ket{0}$, $b^{\dag}_{0,\beta}\ket{0}$ 
and $b^{\dag}_{0,\alpha}b^{\dag}_{0,\beta}\ket{0}$, respectively. Because there cannot exist coherence between different parity eigenstates 
of fermions, the density matrix of the two MZM qubits will always possess the form
\begin{align}
\rho(t)= \begin{pmatrix}
\rho_{00}(t)  &    0    &    0    & \rho_{0d}(t) \\
0        & \rho_{\alpha\alpha}(t) & \rho_{\alpha\beta}(t) & 0 \\
0        & \rho_{\beta\alpha}(t) & \rho_{\beta\beta}(t) & 0 \\
\rho_{d0}(t)  & 0 & 0 & \rho_{dd}(t)
\end{pmatrix}.
\end{align}

At arbitrary time $t$, the relation between the density matrix elements and the Majorana correlation functions reads
\begin{subequations}\label{eq_a_rho}
\begin{align}
&\rho_{00}(t)\!=\!\frac{1}{4}(1\!+\!M_{13}(t)\! +\!M_{24}(t)\!+\!\bar{P}(t))\,,\\
&\rho_{\alpha\alpha}(t)\!=\!\frac{1}{4}(1\!+\!M_{13}(t)\! -\!M_{24}(t)\!-\!\bar{P}(t))\,,\\
&\rho_{\beta\beta}(t)\!=\!\frac{1}{4}(1\!-\!M_{13}(t)\! +\!M_{24}(t)\!-\!\bar{P}(t))\,,\\
&\rho_{dd}(t)\!=\!\frac{1}{4}(1\!-\!M_{13}(t)\! -\!M_{24}(t)\!+\!\bar{P}(t))\,,\\
&\rho_{0d}(t)\!=\! \frac{1}{4}[ -\!M_{14}(t)\! +\!M_{23}(t)\! +\!i \left(M_{12}(t)\! -\!M_{34}(t)\right)]\,,\\
&\rho_{\alpha\beta}(t)\!=\!\frac{1}{4}[ -\!M_{14}(t)\! -\!M_{23}(t)\! -\!i \left(M_{12}(t)\! +\!M_{34}(t)\right)]\,.
\end{align}
\end{subequations}
where $M_{ij}$ ($i,j=1,2,3,4$) is the element of the matrix $\bm{M}(t)$. By substituting Eqs.~\eqref{eq_M} and~\eqref{eq_gamma} 
into Eq.~\eqref{eq_a_rho}, one can obtain the complete solution to the two MZM qubit density matrix at arbitrary time $t$, which is expressed 
in terms of the initial condition of the MZM states and the non-equilibrium Green's functions $\bm{U}(t,t_0)$ and $\bm{V}(t)$. 

Initially, the two MZM qubit density matrix is diagonal and no coherence exists. After the TSC system is coupled to the leads, the 
off-diagonal matrix elements $\rho_{0d}(t)$ or $\rho_{\alpha\beta}(t)$ would, in general, become finite values, i.e., one can generate coherence 
in each MZM qubit state. Moreover, both the dynamical process and the final state can be manipulated by tuning the 
magnetic flux and the coupling strengths. In the following section, we shall discuss the cases of various parameter settings. 
We shall demonstrate that by tuning the magnetic flux $\Phi$, the bias $\mu_L$ and $\mu_R$, and the coupling strengths $\lambda$'s, 
the MZM qubit states can be modified significantly.

\section{Dynamics of the MZMs with various parameter settings}
\label{sec_4}

In this section, we shall study how the coherence dynamics of the MZM qubits are varied under different parameter settings. 
For clarity, in the following analysis, we set the original pairing phase difference $\phi$ of the TSCs to be zero ($\phi=0$) in absence 
of the magnetic flux  [see Eq.~\eqref{eq_delta}], and the initial state of the system as $\rho(t_0)=\ketbra{0}{0}$. 
Firstly, we investigate the general dynamics of two MZM qubits with different parities. For simplicity, we avoid the complicated tunneling effects 
due to the structure of the leads and simply take the wide-band limit of the spectral density functions: 
$[\bm{J}^{0}_{L/R}]_{\alpha\alpha}\!=\![\bm{J}^{0}_{L/R}]_{\alpha\beta}\!=\![\bm{J}^{0}_{L/R}]_{\beta\alpha}\!=\![\bm{J}^{0}_{L/R}]_{\beta\beta}\!=\!\Gamma_{0}$ 
with $\Gamma_0$ standing for a constant. The matrix elements of the retarded Green's functions are then explicitly given by 
\begin{subequations}
\begin{align}
    [\bm{U}_{L/R}]&_{\alpha\alpha}(t,t_0)\!=\![\bm{U}_{L/R}]_{\beta\beta}(t,t_0) \notag\\
    \!=&\frac{1}{2}\Big[e^{-\Gamma_{0}(1+y)(t-t_0)}
    \!+\!e^{-\Gamma_{0}(1-y)(t-t_0)}\Big], \label{U_a}\\
    [\bm{U}_{L/R}]&_{\alpha\beta}(t,t_0)\!=\![\bm{U}_{L/R}]_{\beta\alpha}(t,t_0) \notag\\
    \!=&\frac{1}{2}\Big[e^{-\Gamma_{0}(1+y)(t-t_0)}
    \!-\!e^{-\Gamma_{0}(1-y)(t-t_0)}\Big] \label{U_b},
\end{align}
\end{subequations}
where $y\!=\!\cos{[\pi\Phi/\Phi_0]}$. Note that $\bm{U}(t,t_0)$ and thus the MZM qubit density matrix $\rho(t)$ show $2\Phi_0$-periodicity as a function 
of the magnetic flux $\Phi$. The diagonal elements of $\bm{U}(t,t_0)$ describe the decays of the MZM qubits. 
From Eq.~(\ref{U_a}), one can see that the decay of MZM qubit states consist of two parts with different decay times, namely 
$[\Gamma_{0}(1+y)]^{-1}$ and $[\Gamma_{0}(1-y)]^{-1}$ respectively. Therefore, if $y\neq \pm1$, i.e., $\Phi/\Phi_0\neq n$ ($n$ stands 
for an integer), the MZM qubits will inevitably decay away. 
Furthermore, it is obvious from Eq.~(\ref{U_a}) that for $y\!=\!\pm 1$, i.e., $\Phi/\Phi_0=n$, there exist dissipationless modes and part of the 
MZM qubit states will not decay. For the parameters considered above, when $\Phi/\Phi_0$ 
is an even integer, 
the system generates two dissipationless MZM modes reading $\frac{1}{2}(\gamma_{\alpha L}-\gamma_{\beta L})$ and $\frac{1}{2}(\gamma_{\alpha R}-\gamma_{\beta R})$, 
while for $\Phi/\Phi_0$ being an odd integer, 
the system forms two dissipationless modes reading $\frac{1}{2}(\gamma_{\alpha L}\!+\!\gamma_{\beta L})$ and $\frac{1}{2}(\gamma_{\alpha R}\!+\!\gamma_{\beta R})$. 
\begin{figure}[]
\includegraphics[width=8.6cm]{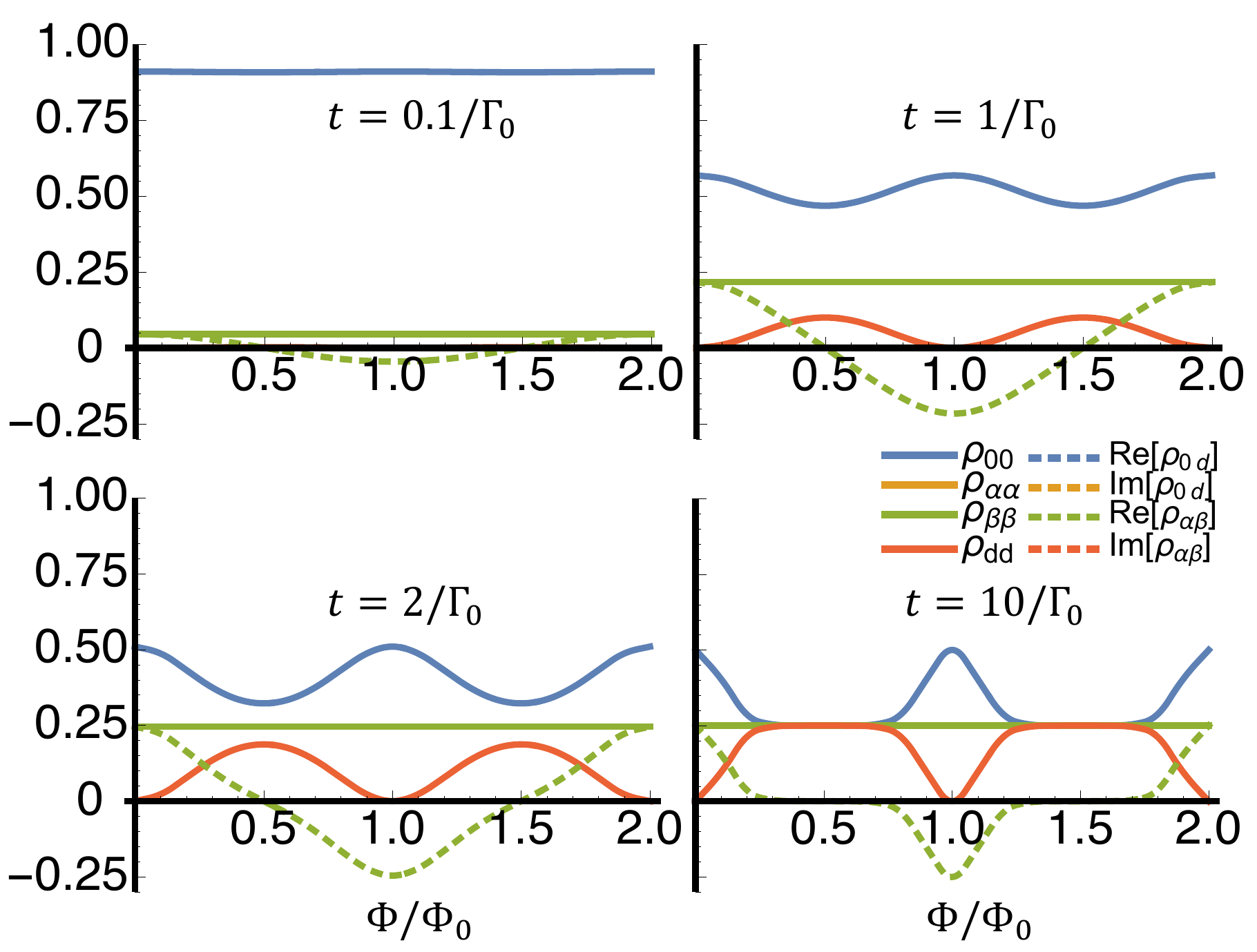}
\caption{\small (Color Online) The density matrix elements of the MZM states with varying magnetic flux at time (a) (top left) 
$t=0.1/\Gamma_0$, (b) (top right) $t=1/\Gamma_0$, (c) (bottom left) $t=2/\Gamma$ and (d) (bottom right) $t=10/\Gamma_0$. 
The other parameters of the  device are $T_L\!=\!T_R\!=\!0.1\hbar\Gamma_0/k_\text{B}$, $\mu_L\!=\!\mu_R\!=\!0$, and 
$\bm{J}^{0}_{L}/\Gamma_{0}\!=\!
\bm{J}^{0}_{R}/\Gamma_{0}\!=\!\mqty*(1 & 1 \\ 1  &  1)$, and the initial state of the system is $\rho(t_0)\!=\!\ketbra{0}{0}$. 
It is noteworthy that since the evolution of $\rho_{\alpha\alpha}$ and $\rho_{\beta\beta}$ are precisely the same, their curves 
coincide with each other.} \label{fig_phi}
\end{figure}

On the other hand, the off-diagonal elements of $\bm{U}(t,t_0)$ characterize the correlations between MZMs $\gamma_{\alpha L(R)}$ 
and $\gamma_{\beta L(R)}$, and hence relate to the MZM qubit coherence. One can see from Eq.~(\ref{U_b}) that 
$[\bm{U}_{L/R}]_{\alpha\beta}$ increases from zero initially, implying that the correlations between MZMs are building up. 
If there are no dissipationless modes, these build-up MZM correlations will eventually vanish. When dissipationless MZM
modes exist, the MZM correlation functions $[\bm{U}_{L/R}]_{\alpha\beta}$ will reach a steady value of $1/2$ (for 
$\Phi/\Phi_0$ being odd) or $-1/2$ (for  
$\Phi/\Phi_0$ being even). 
To study the MZM qubit dynamics, the evolution of the density matrix elements of the MZM states is shown in Fig.~\ref{fig_phi}. 
In the case of zero bias, i.e. $\mu_L=\mu_R=0$, the  MZM qubit will eventually decay to a maximally mixed 
state if there is no dissipationless MZM mode. As mentioned above, the qubit coherence (described by the off-diagonal 
elements of the density matrix) grows from zero initially and fades away as the MZMs decay. Explicitly, ${\rm Re}[\rho_{\alpha\beta}]$ 
quickly grows from zero to $0.25$ within $t\sim 1/\Gamma_0$ (see Fig.~\ref{fig_phi}b), then decreases at a decay rate depending 
on the magnetic flux $\Phi/\Phi_0$ (see Eq.~\eqref{U_b}). On the other hand, the dissipationless MZM mode, which exists when 
$\Phi/\Phi_0=0$, $1$ or $2$, will preserve part of the initial qubit state information and keep the MZM qubits away from 
a maximally mixed state. Note that two different dissipationless MZM modes are formed at $\Phi/\Phi_0=0$ and $\Phi/\Phi_0=1$ 
(see ${\rm Re}[\rho_{\alpha\beta}]$ in Fig.~\ref{fig_phi}), showing the $2\Phi_0$-periodicity of the MZM qubit states.

\begin{figure}[!htb]
\includegraphics[width=8.9cm]{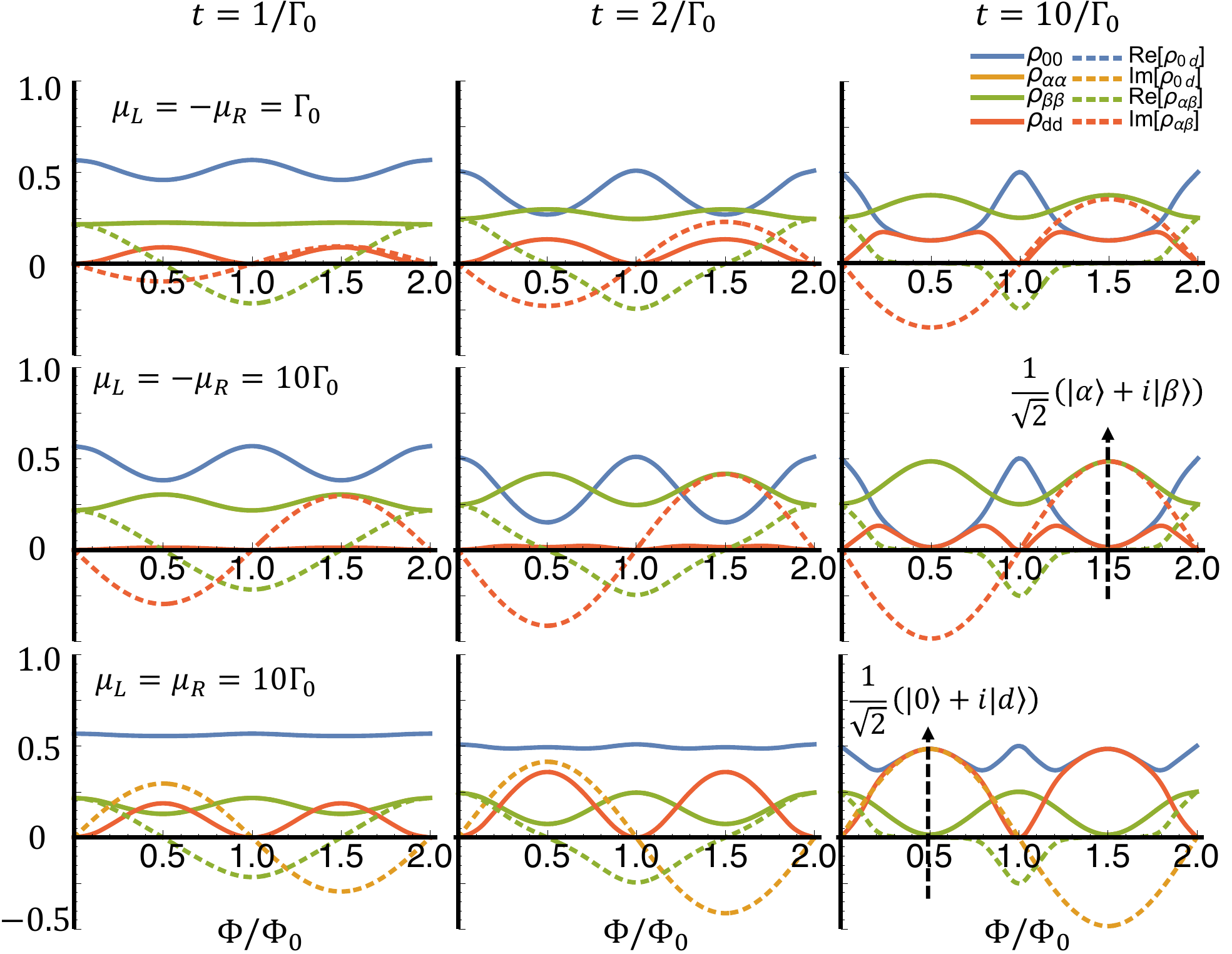}
\caption{\small (Color Online) The MZM qubit state evolution with various magnetic flux and bias voltages. The common conditions for 
the device are $T_L\!=\!T_R\!=\!0.1\hbar\Gamma_0/k_{\text{B}}$, $\bm{J}^{0}_{L}/\Gamma_{0}\!=\!
\bm{J}^{0}_{R}/\Gamma_{0}\!=\!\mqty*(1 & 1 \\ 1  &  1)$, and the initial states of the system are all set to be $\rho(t_0)=\ketbra{0}{0}$. 
It is noteworthy that the curves of $\rho_{\alpha\alpha}$ and $\rho_{\beta\beta}$ coincide with each other. The left, middle and right 
columns show the MZM density matrix elements at time $t=1/\Gamma_0$, $2/\Gamma_0$ and $10/\Gamma_0$, respectively. While 
the rows show the MZM density matrix elements with the applied bias $\mu_L=-\mu_R=\hbar\Gamma_0$ (top row), $\mu_L=-\mu_R=10\hbar\Gamma_0$ 
(middle row) and $\mu_L=\mu_R=10\hbar\Gamma_0$ (bottom row) respectively. The dashed arrows are marked to show the state 
$(\ket{\alpha}\!+\!i \ket{\beta})/\sqrt{2}$ and $(\ket{0}\!+\!i \ket{d})/\sqrt{2}$.} \label{mu}
\end{figure}

Next, the bias $\mu_L$ and $\mu_R$ can be tuned so that two qubit steady states will not become a maximally mixed state. In this case, 
apart from the mere damping of the MZMs, electrons and holes can be pumped into or out of the two TSCs from the leads. As a result, 
the even and odd parity qubit states are not equally occupied. Furthermore, MZM qubit coherence with definite parity can also be generated by applying bias 
(See the curves corresponding to ${\rm Im}[\rho_{\alpha\beta}]$ and ${\rm Im}[\rho_{0d}]$ in Fig.~\ref{mu}). If the bias $\mu_L$ and $\mu_R$ 
are large enough, MZM qubits can evolve to a state with almost definite parity and perfect coherence. For instance, in the case of 
$\Phi/\Phi_0\!=\!\frac{3}{2}$, a large anti-symmetric bias (e.g. $\mu_L\!=\!-\mu_R\!=\!10\hbar\Gamma_{0}$) leads the system to the almost pure 
qubit state with odd parity, namely, $(\ket{\alpha}\!+\!i\ket{\beta})/\sqrt{2}$. While in the case of $\Phi/\Phi_0\!=\!\frac{1}{2}$, a large 
symmetric bias (e.g. $\mu_L\!=\!\mu_R\!=\!10\hbar\Gamma_{0}$) leads the system to to the almost pure qubit state with even parity, 
namely, $(\ket{0}\!+\!i \ket{d})/\sqrt{2}$.

\begin{figure}[!htb]
\includegraphics[width=8.9cm]{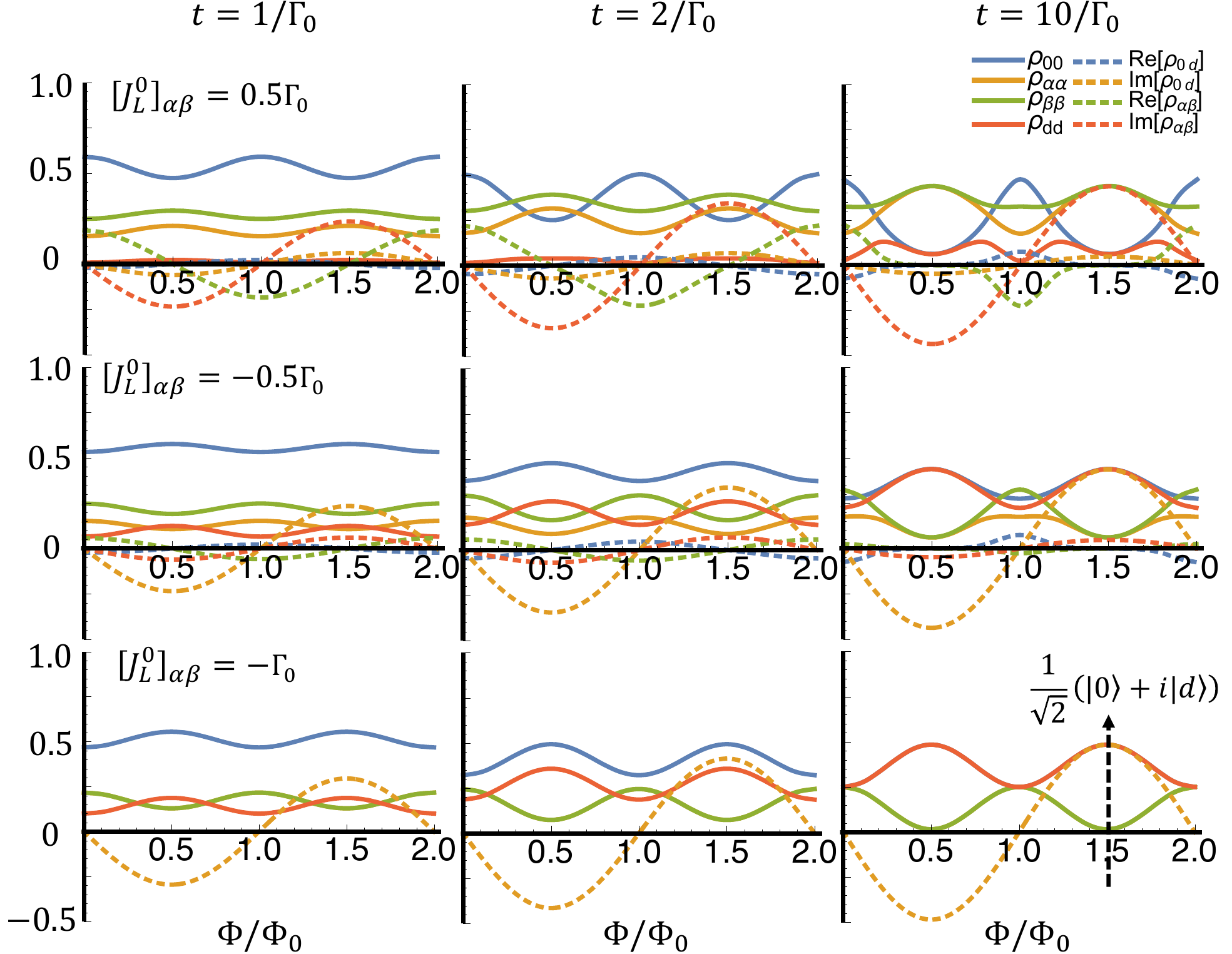}
\caption{\small (Color Online) The MZM qubit state evolution with various magnetic flux and cross-coupling strength. 
The common conditions for the device are set to be $T_L=T_R=0.1\hbar\Gamma_0/k_{\text{B}}$, $\mu_L=-\mu_R=10\hbar\Gamma_0$ 
and the initial states of the system are all set as $\rho(t_0)=\ketbra{0}{0}$. It is noteworthy that the evolution of 
$\rho_{\alpha\alpha}$ and $\rho_{\beta\beta}$ are precisely the same so their curves coincide with each other. 
The left, middle and right columns show the MZM density matrix elements at time $1/\Gamma_0$, $2/\Gamma_0$ and 
$10\Gamma_0$ respectively. While the rows show the MZM density matrix elements with cross-coupling strength 
$[\bm{J}^{0}_{L}]_{\alpha\beta}=0.5\Gamma_0$ (top row), $[\bm{J}^{0}_{L}]_{\alpha\beta}=-0.5\Gamma_0$ (middle row) 
and $[\bm{J}^{0}_{L}]_{\alpha\beta}=-\Gamma_0$ (bottom row) respectively. The dashed arrows are marked to show the state 
$(\ket{0}\!+\!i \ket{d})/\sqrt{2}$.} \label{fig_LR}
\end{figure}

\begin{figure}[!htb]
\includegraphics[width=5cm]{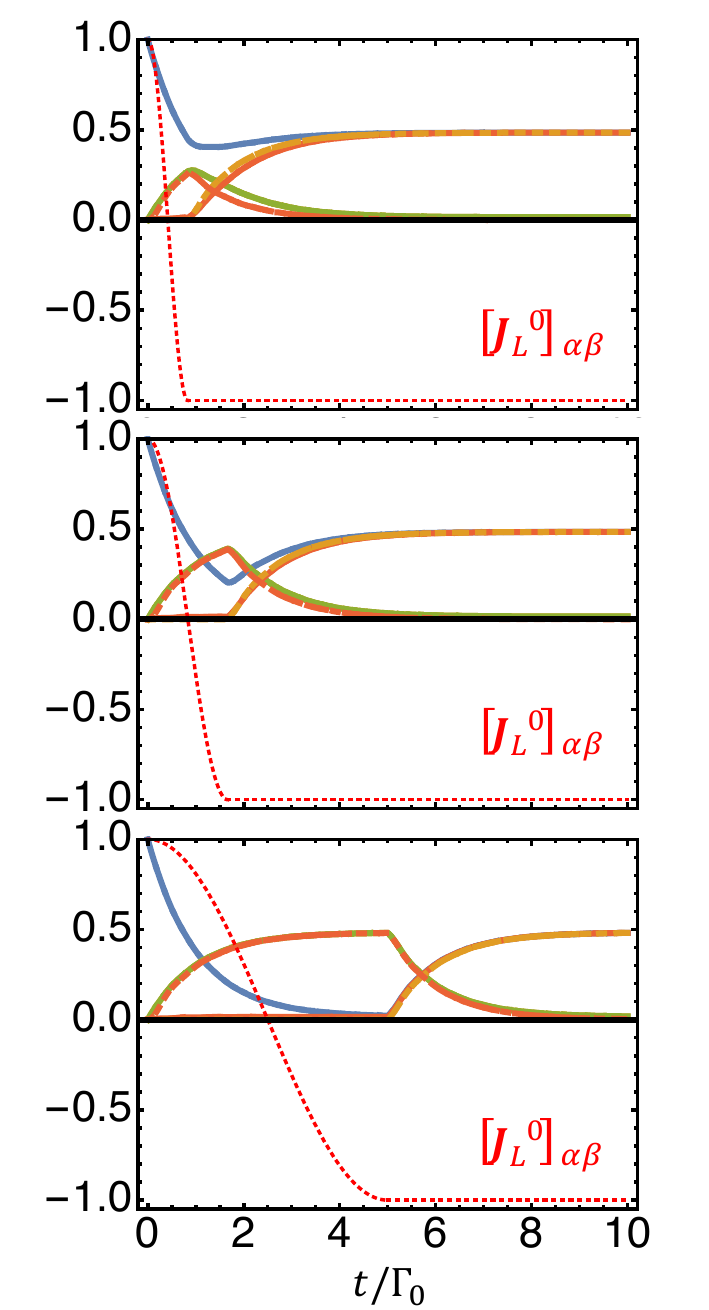}
\caption{\small (Color Online) The cross coupling strength $[\bm{J}^{0}_{L}]_{\alpha\beta}/J_{0}$ (red dotted line) is tuned at different rates. The density matrix elements are plotted with the same line patterns as in the previous figures. (a) (top) The MZMs relaxes directly to the even parity state. (b) (middle) The MZMs relax partially to the odd parity state then ``turns'' its 
relaxation to the even parity state. (c) (bottom) The MZMs relaxes fully to the odd parity state then relaxes to the even parity state.} \label{fig_parityflip}
\end{figure}

As we demonstrated above, the applied bias leads to the polarization of the MZM state parity. Actually, this parity polarization can also 
be controlled by tuning the lead-TSC couplings. Specifically, when a bias of $\mu_L=-\mu_R=10\hbar\Gamma_0$ is applied, the dominated parity 
is flipped when the cross-coupling strength  $[\bm{J}^{0}_{L}]_{\alpha\beta}/\Gamma_{0}$ changes from positive to negative 
(see Fig.~\ref{fig_LR}). We demonstrate in detail this parity-flip dynamics in Fig.~\ref{fig_parityflip}, in which the cross-coupling strength 
of the left-hand side $[\bm{J}^{0}_{L}]_{\alpha\beta}/\Gamma_{0}$ is tuned so that it changes from $1$ to $-1$ at different rates. 
Note that we have fixed the coupling strengths between the TSCs and the right lead. When $[\bm{J}^{0}_{L}]_{\alpha\beta}/\Gamma_{0}$ 
is tuned within a very short time ($\sim 1/\Gamma_{0}$), the MZMs will relax directly to the even parity state $(\ket{0}\!+\!i \ket{d})/\sqrt{2}$ 
(see Fig.~\ref{fig_parityflip}a). When the changing time of $[\bm{J}^{0}_{L}]_{\alpha\beta}/\Gamma_{0}$ becomes a little longer 
($\sim 2/\Gamma_{0}$), as shown in Fig.~\ref{fig_parityflip}b, the MZMs will relax partially to the odd parity state but then ``turn'' its 
relaxation to the even parity state. This is because the coupling changes so fast that the MZM states cannot reach full relaxation. Finally, 
when the changing time of $[\bm{J}^{0}_{L}]_{\alpha\beta}/\Gamma_{0}$ is long enough ($\sim 5/\Gamma_{0}$), the MZMs 
relaxes from the initial state $\ketbra{0}{0}$ to the odd parity state $(\ket{\alpha}\!+\!i\ket{\beta})/\sqrt{2}$, then relaxes again 
to the odd parity state, which is a parity flip between two MZM qubit states [see Fig.~\ref{fig_parityflip}c].

\section{Conclusion}
\label{sec_5}

In this paper, we propose a Majorana Aharonov-Bohm interferometer to control the MZM states. In this device, electrons and holes transport from 
one lead to another through the rectangular ring formed by the four spatially separated MZMs. Through this transport process, the qubit states of the MZMs 
evolves and manifests various features such that their state evolution can be tuned by setting the parameters of the interferometer. 

With path-integral approach in the coherent state representation, we obtain the exact master equation of the two MZM qubits, one qubit has the even 
fermion parity and the other has old parity. The effects of the leads 
on the system are clearly revealed in the structure of the master equation. Formally,  the MZMs in the left and right evolve independently, 
which are respectively influenced by the leads on the left and right side. However, because the renormalized Hamiltonian and damping 
coefficients all depend on the global quantity, i.e., the total magnetic flux $\Phi$, the qubit state evolution of the MZMs actually involves interference 
effect. Note that the density matrix of two MZM qubits shows $2\Phi_0$-periodicity of the magnetic flux. 

It is shown that by tuning the magnetic flux, the bias voltage of the leads, and the TSC-lead coupling strength, the interference property of the 
MZM qubit states can be modified significantly. The two decoherence rates $\Gamma_{0}(1+y)$ and $\Gamma_{0}(1-y)$ can be changed 
by tuning the magnetic flux, and dissipationless modes can be formed for certain values of the magnetic flux. By setting bias among the 
leads and the TSCs,  MZM qubit states can be drawn away from approaching the maximally mixed state. The fermion parity of the MZM 
qubit can be polarized and the MZM qubit coherence can also be generated. The parity of the target state can be controlled by 
setting the bias voltages in a particular configuration, or by tuning the TSC-lead coupling through the controlling gates. If the bias is large 
enough, the state can evolve to a nearly pure coherent MZM qubit state within the same parity. Moreover, the switch between different parity 
qubit states can be realized by changing the cross-coupling strength from positive (negative) to negative (positive) at suitable rates.

\acknowledgments
We thank Lian-Ao Wu and Yu-Wei Huang for helpful discussions. This work is supported by the Ministry of Science and Technology of the 
Taiwan under the Contracts No. MOST-108-2112-M-006-009-MY3.

\bibliographystyle{apsrev4-1}
\bibliography{references}

\end{document}